\documentclass[sn-mathphys-num,Numbered]{sn-jnl}%

\usepackage{graphicx}
\usepackage{dcolumn}
\usepackage{bm}
\usepackage{mathtools}
\usepackage{mathrsfs, amsmath, wasysym}
\usepackage[mathscr]{eucal}
\usepackage{xfrac}
\usepackage{graphicx}
\usepackage{dcolumn}
\usepackage{bm}
\usepackage{color}
\usepackage{tabularx}
\usepackage{hyperref}
\hypersetup{colorlinks,linkcolor=blue,citecolor=blue,urlcolor=blue}
\usepackage[normalem]{ulem}
\usepackage[all]{hypcap}
\usepackage[dvipsnames,usenames,table]{xcolor}
\usepackage{chngcntr}
\usepackage{amsmath, amssymb}
\usepackage{chemformula}

\newcommand{\AuCu}{Au$_x$Cu$_{1-x}$}

\newcommand{\bk}{{\bf k}}

\newcommand{\bx}{{\bf x}}

\DeclareMathOperator{\EX}{\mathbb{E}}

\begin{document}

\title[Towards Accurate Prediction of Configurational Disorder Properties in Materials using Graph Neural Networks]{Towards Accurate Prediction of Configurational Disorder Properties in Materials using Graph Neural Networks}

\author*[1]{\fnm{Zhenyao} \sur{Fang}}\email{z.fang@northeastern.edu}

\author*[1]{\fnm{Qimin} \sur{Yan}}\email{q.yan@northeastern.edu}

\affil[1]{\orgdiv{Department of Physics}, \orgname{Northeastern University}, \city{Boston}, \postcode{02115}, \state{Massachusetts}, \country{United States}}

\abstract{
The prediction of configurational disorder properties, such as configurational entropy and order-disorder phase transition temperature, of compound materials relies on efficient and accurate evaluations of configurational energies. Previous cluster expansion methods are not applicable to configurationally-complex material systems, including those with atomic distortions and long-range orders. In this work, we propose to leverage the versatile expressive capabilities of graph neural networks (GNNs) for efficient evaluations of configurational energies and present a workflow combining attention-based GNNs and Monte Carlo simulations to calculate the disorder properties. Using the dataset of face-centered tetragonal gold copper without and with local atomic distortions as an example, we demonstrate that the proposed data-driven framework enables the prediction of phase transition temperatures close to experimental values. We also elucidate that the variance of the energy deviations among configurations controls the prediction accuracy of disorder properties and can be used as the target loss function when training and selecting the GNN models. The work serves as a fundamental step toward a new data-driven paradigm for the accelerated design of configurationally-complex functional material systems.
}


\maketitle

\section{Introduction} \label{sec:intro}
Disordered materials have attracted much attention in the community in recent years due to their exotic structural and electronic properties such as Anderson localization and Mott-like conduction \cite{Anderson58p1492, Cutler69A1336, Pu22p116801}, novel phonon scattering channels and lattice dynamics \cite{Wang20p155, Alam21p104202, Huang21p073901}, enhanced ductility and mechanical strength over a wide temperature range in high-entropy alloys \cite{Jiang22HEA, Zaddach13HEA, Gludovatz14HEA}, and regulated electronic states and atomic sites for catalysts \cite{Sharma21p2552, Xie21p1738}, endowing them with promising applications in electronic devices and energy materials. Depending on its chemical nature, disorder effects in materials can be classified into structural disorder characterized by disrupted chemical bonding network (such as vacancies, dislocations, and dangling bonds) \cite{Cliffe10p125501}, and configurational (compositional) disorder characterized by crystallographic sites being occupied by irregular atomic species \cite{Cordell21p024604}. In this work, we focus on the latter type of disorder.

To numerically access the configurational disorder properties, such as the order-disorder phase transition temperature \cite{Yang20p085402, Reese69p905} and the configurational entropy \cite{Senkov15p6529, Oses20p295}, Monte Carlo (MC) simulations are often carried out with Metropolis sampling \cite{Prokhorenko18p80} or Wang-Landau sampling \cite{Wang01p2050, Zhou06p120201}. With the Metropolis sampling, the free energy is obtained by performing thermodynamic integration numerically using the average energies at each temperature. As for the Wang-Landau sampling, the density of states, instead of the average energies, is estimated, and the configurational entropy and the heat capacity at any arbitrary temperature can then be evaluated. However, both sampling methods require evaluating a large number of supercell configurational energies efficiently and accurately in order to achieve convergence, thus not applicable to first-principles methods especially when the cell size is very large.

One commonly used approach to this problem is the cluster expansion (CE) method \cite{Sanchez93p14013, Drautz19p014104, Seko09p165122, Cao18p2401, Chang19p325901, Barroso22p4504}, where the cell is decomposed into different atomic clusters up to a cutoff size, and the total energy of the cell is expanded into the effective cluster interactions of these clusters. This method has been applied to evaluate the total energies of different configurations and further to calculate the disorder properties effectively. However, it suffers from several limitations. Due to the limited cluster size in practice, it cannot capture long-range orders in materials, which may affect the phase stability and electronic structures \cite{Laks92p12587}. Besides, the definitions of clusters strongly depend on the atomic positions, restricting this method from adapting to lattice distortions and local atomic displacements induced by atomic relaxations or thermal effects \cite{Kadkhodaei21p3326}.

Since 2018, graph neural networks (GNNs) have been applied to studying the structure-property correlation in complex solid-state materials. Instead of relying on human-selected descriptors, GNNs can autonomously learn latent representations of materials and make fast and accurate atom-, bond-, and material-level predictions. \cite{Chen19GNN, Reiser22GNN, Fung21GNN} Therefore, in this work, we propose to employ GNNs to evaluate the configurational energies and to access the disorder properties in configurationally-complex compound materials accurately, because of their high representation capability and versatile adaptability to realistic simulation scenarios, including lattice distortions, atomic displacements, and various types of defects \cite{Reiser22GNN, Fung21GNN}. Especially, we construct attention-based GNN models from Transformer neural networks \cite{Shi21Transformer}, leveraging masked self-attentional layers to obtain configurational energies efficiently. The model is trained on the face-centered tetragonal (fct) gold copper (\AuCu) alloy dataset obtained from density functional theory (DFT) calculations. The trained model can well reproduce the DFT configurational energies, with a mean absolute error (MAE) of 2.76 meV/atom, which eventually leads to the prediction of order-disorder phase transition temperature that is comparable to experimental observation. When random atomic displacements are introduced, which is beyond the capability of the CE method, the GNN model can still evaluate the configurational energies accurately (with MAE being 5.02 meV/atom). The predicted phase transition temperature is slightly lower than the undistorted case, suggesting structural disorder can enhance the configurational disorder. Furthermore, we reveal the connections between the variance of the energy deviations among configurations and the accuracy of configurational entropy and heat capacity predictions, providing new guidance on future data-driven studies of the configurational disorder properties in materials. 

\begin{figure}[htb]
\centering
\includegraphics[width=\linewidth]{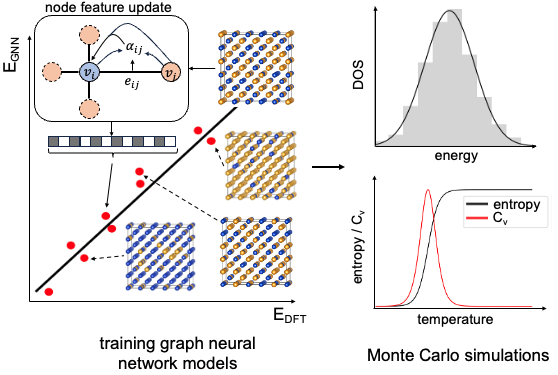}
\caption{A schematic plot of the workflow to use GNN to calculate the disorder properties of compound materials. The input crystal structure is converted to a graph, whose nodes and edges represent the atomic species and the interatomic distances, respectively. Through several attention-based convolution layers, where the node features aggregate and interact with each other, global features are extracted and further processed by linear layers to predict the energy. The well-trained GNN model is subsequently utilized in Monte Carlo simulations to acquire the final disorder-related properties, such as the configurational density of states, configurational entropy, and heat capacity.}
\label{fig:workflow}
\end{figure}

\section{Results} \label{sec:results}
\subsection{Pristine \AuCu~Structure Dataset} \label{sec:pristine}
\begin{figure}[htb]
\centering
\includegraphics[width=\linewidth]{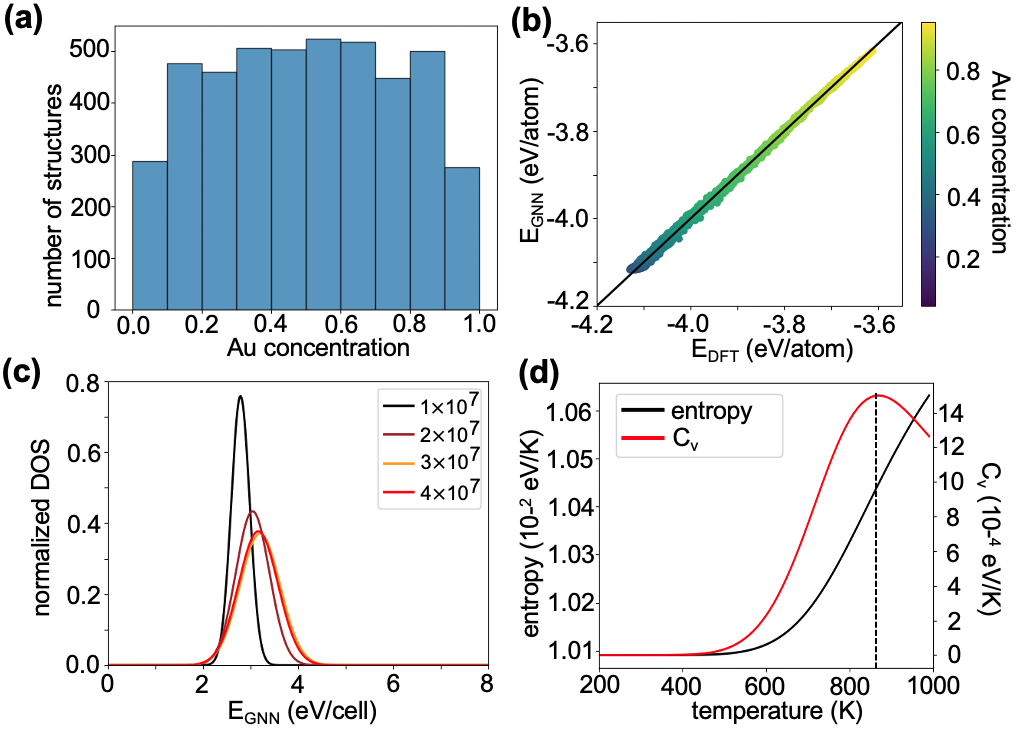}
\caption{The performance of the Transformer-based GNN model on the pristine \AuCu~structure dataset. (a) The histogram of the number of configurations for each Au concentration. (b) The comparison between the configurational energies (per atom) predicted from DFT ($E_\text{DFT}$) and the optimal GNN model ($E_\text{GNN}$). The color of each point represents the concentration $x$ of Au atoms in the configuration. (c) The calculated normalized density of states of pristine AuCu during the MC simulation process (per cell). The total number of MC steps is $4 \times 10^7$. (d) The calculated configurational entropy (black curve) and the heat capacity (red curve) at different temperatures.}.
\label{fig:pristine}
\end{figure}

We choose fct AuCu to construct our dataset. Although both experimental and numerical results on the configurational entropy and the phase transition temperature (683~K) were reported \cite{Chang19p325901, Wei87p4163}, those results are based on the face-centered cubic (fcc) structure, which is higher in energy than the ground-state fct structure by 0.016~eV per formula unit and can only be stable under pressure. However, these two structures differ only in the lattice constants ($\Delta a / a = 6.9\%$ and $\Delta c / c = 15\%$), and thus we expect their disorder properties, especially the phase transition temperature, to be similar. Using first-principles calculations, we construct our dataset for pristine \AuCu~containing 4500 configurations in a $5 \times 5 \times 4$ supercell with 200 atoms, covering all chemical compositions $x \in [0, 1]$. The concentration distribution of the dataset is shown in Fig.~\ref{fig:pristine}(a).

The MAE of the best GNN model built upon Transformer convolution layers on the testing set of our fct \AuCu~dataset is 2.76~meV/atom, similar to previously reported cluster expansion methods on fcc-structure AuCu (4.49~meV/atom) \cite{Chang19p325901}. The comparison between the predicted energy and the DFT energy is shown in Fig.~\ref{fig:pristine}(b), indicating the capability of our attention-based GNN model to evaluate the configurational energies accurately for all compositions of \AuCu. 

We compare this result with other commonly used GNNs with attention mechanism, such as the graph attention network \cite{GAT1, GAT2}, and those without attention mechanism, including the crystal graph neural network (CGNN) \cite{Xie18p145301} and the edge-conditioned neural network (ECNN) \cite{Simonovskuy17ECConv, Gilmer17ECConv} (details about these convolution layers can be found in Supplementary Note S1). The MAEs of the best GNN model built for each type of convolution layers are summarized in Tab.~\ref{tab:conv_layer}, suggesting that GNN models with the attention mechanism perform better than those lacking this mechanism when evaluating configurational energies. In attention-based neural networks, node features are updated by summing up adjacent node features weighted by the attention coefficients, calculated from the node features and the edge features and thus containing the information on the similarity between the two nodes. Specifically in Transformer convolution layers, the central node and the adjacent nodes are considered as queries and keys respectively, and the overlap between the queries and the keys reflects the similarity between the two nodes (atoms). This attention mechanism updates the central node features by a linear combination of adjacent node features weighted by the overlap between the query vector and key vectors in the latent space, thus allowing the network to effectively capture the chemical distinctions between neighboring atoms and increasing its efficacy in predicting properties related to crystal structure. Besides, since the attention coefficients depend on the node features, they are dynamic and can vary across different convolution layers, as compared to the static weight matrices that are fixed across the convolution layers in models without the attention mechanism. Therefore, we expect that in general models with the attention mechanism are better than those without it.  Finally, we also observe the superior performance of the ECNN model compared to the CGNN model, possibly because the ECNN model adopts a multilayer perceptron layer to process edge features while the CGNN model employs only a linear transformation on edge features. This distinction makes the former more adaptable in evaluating configurational energies.

\begin{table*}[htb]
    \centering
    \begin{tabular}{|c|c|c|c|c|}
    \hline
    & GAT & Transformer & CGNN & ECNN \\ \hline
    MAE & 2.81 & 2.76 & 8.14 & 6.25 \\ \hline
    variance & 13.53 & 13.20 & 78.19 & 40.90 \\ \hline
    \end{tabular}
    \caption{The MAE (meV/atom) and the variance (meV\textsuperscript{2}/atom) on the pristine \AuCu~datasets using different types of graph neural network models.}
    \label{tab:conv_layer}
\end{table*}

We then apply the optimal GNN model based on Transformer attention mechanism to MC simulations to obtain the configurational properties of stoichiometric AuCu systems. Because the configurational space is complicated for such large supercells, we choose a bin size of 0.2 eV in the Wang-Landau sampling method to expedite convergence. Since the MAE of the trained GNN model is around 2.76~meV/atom, we anticipate that each configuration has an energy deviation of 0.6~eV/cell on average, justifying our choice of bin size. With this choice of the bin size, the MC simulation takes around $4 \times 10^7$ steps to converge, beyond the capability of DFT methods. We show the intermediate and final density of states that are fitted to a Gaussian distribution in Fig.~\ref{fig:pristine}(c). The density of states is normalized such that the sum of the density of states equals to one; while the normalization constant is the total number of configurations of stoichiometric AuCu in a 200-atom supercell, i.e., $\Omega_\text{max} = \binom{100}{200}$. The peak of the density of states is higher in energy than the ground state by 3.2~eV/cell. From the density of states, we can calculate the configurational entropy and the heat capacity according to Eq.~\ref{eqn:thermodynamic}, shown in Fig.~\ref{fig:pristine}(d). The position of the heat capacity peak, indicating the order-disorder phase transition temperature, is at 870~K (for fct AuCu), similar to the experimental values. As temperature further increases, the configurational entropy gradually approaches the theoretical limit for the fully disordered phase (within a 200-atom supercell) $S_\text{max} = k_B \ln{\Omega_\text{max}} = 1.17 \times 10^{-2}$~eV/K.

We also note that while training and selecting the best GNN model for MC simulations, batch normalization features can affect the predictions of disorder properties significantly and should be disallowed. When training GNN models, configurations are grouped into batches of a reasonable size to accelerate the training process. In general, too small batches can lead to instability of gradients and optimization process, while too large batches can lead to huge memory requirements and possible overfitting. However, in MC simulations, energy evaluations are performed on a single configuration consecutively. When using trained models with batch normalized features to predict the energy of one configuration, incorrect predictions may arise because in this case the batch mean and batch variance are highly influenced by the specific configuration in the batch, but not reflecting the distribution of the features of the whole dataset. In Supplementary Note S2, we show benchmark calculations for the GNN model with batch normalization features. By evaluating the configurations in batches (containing 58 configurations), the MAE of the best model is 3.52 meV/atom, but when evaluated consecutively, the MAE loss for the model increases drastically to 2.76~eV/atom, and the corresponding prediction on the order-disorder phase transition is incorrect (183~K).

\subsection{Distorted \AuCu~Structure Dataset} \label{sec:distorted}
To showcase the adaptability of GNN models to realistic simulation scenarios, such as atomic relaxations where cluster expansion methods face convergence challenges, we constructed another dataset containing 4500 configurations with random atomic displacements. The maximum amount of displacement for each atom is chosen to be 0.35~\AA. 

\begin{figure}[htb]
\centering
\includegraphics[width=\linewidth]{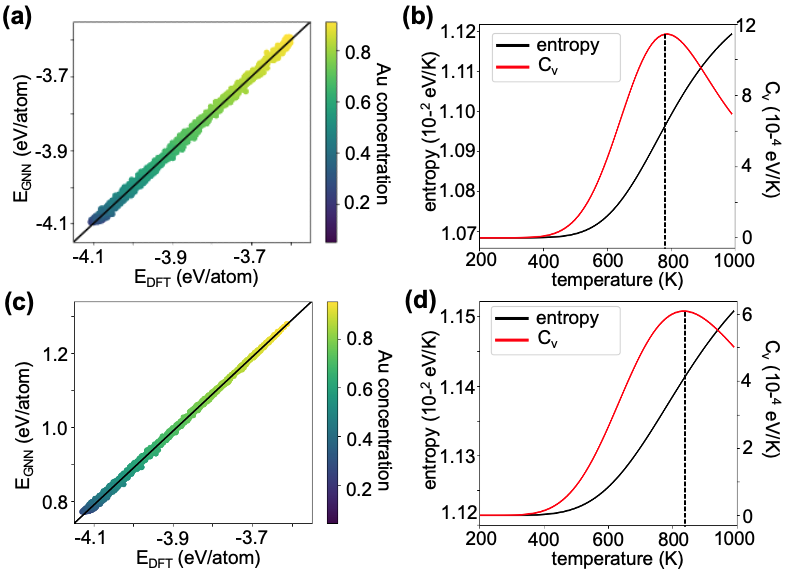}
\caption{The performance on the dataset with random atomic displacements, using MSE as the target loss function. (a) The comparison between the configurational energies (per atom) obtained from DFT ($E_\text{DFT}$) and the optimal GNN model ($E_\text{GNN}$). (b) The calculated configurational entropy (black curve) and the heat capacity (red curve) at different temperatures. (c, d) Same as (a, b), but on a dataset without displacements and using variance as the target loss function.}.
\label{fig:distortion_variance}
\end{figure}

The MAE on the testing set of this dataset containing distorted structures is 6.43~meV/atom, and the comparison between the DFT and GNN energies is shown in Fig.~\ref{fig:distortion_variance}(a). Larger deviations from DFT energies are primarily observed in configurations with lower Au concentration. We further apply the trained model to calculate the disorder properties, shown in Fig.~\ref{fig:distortion_variance}(b). We generate one supercell with random atomic displacements and apply it for MC simulations while keeping all the atomic positions fixed. The calculated phase transition temperature is 788~K, lower than that for the pristine structure, suggesting that structural disorder can enhance the configurational disorder; the introduction of structural disorder increases the likelihood of the material being in the configurational disordered phase.

A potential avenue for enhancing the performance of GNN models for energy predictions of disordered systems with distorted structures lies in the modification of how crystal structures are represented in the form of graphs. In our current method, the nodes represent the atomic species and the edges contain only bond distance information. But other local chemical information can also be included in the graphs, such as the directional information (for instance, bond directions expanded in spherical harmonic coefficients) and the multiplet interactions that can be included in hypergraphs \cite{Feng19p3558} or higher-order graphs \cite{Morris18p4602}. These graph structures may capture the much more complex chemical environment of each atom after introducing random atomic displacements. An alternative avenue for improvement is to modify the network structure. For example, the Au concentration can be treated as a global feature for each configuration. This feature could be concatenated with the global features from the nodes and jointly passed through the linear layers to determine the total energy of the configuration. Despite room for these improvements, our current model already suffices for the effective predictions of disorder properties of complex alloy systems with distorted structures.

\section{Discussions} \label{sec:variance}
Next, we would like to draw the reader's attention to the correlation between disorder properties and the optimization process of GNN models. From Eq.~\ref{eqn:thermodynamic}, it can be shown that neither the heat capacity nor the configurational entropy are affected by an overall energy shift $\Delta E$ to all configurations (see Supplementary Note S3). Based on this observation, we postulate that when training and selecting the best GNN model to predict configurational entropy and heat capacity, the variance of the differences between $E_\text{DFT}$ and $E_\text{GNN}$ among all configurations, instead of the mean of the energy differences, controls the accuracy of predictions and thus can be chosen as the target quantity for optimization.

In the following, we denote $\hat{E}_i$ as the energy predicted by GNN models and $E_i$ as the "ground-truth" energy obtained by DFT for configuration $i$, and their difference as $e_i = \hat{E}_i - E_i$ (in the following, thermodynamic quantities with hats are calculated from $\hat{E}_i$, and those without hats are from $E_i$). We assume that those energy differences are independent and identically distributed random variables, following the normal distribution $\mathcal{N}(\mu, \sigma^2)$, with mean $\mu$ and variance $\sigma^2$; thus we have $\EX[e_i] = \mu$, $\EX[e_i^2] = \mu^2 + \sigma^2$, and $\EX[e_i e_j] = \mu^2$ for all $i$ and $j \neq i$. 

To derive the expected deviations of configurational entropy and the heat capacity due to $e_i$'s, we first focus on the difference of the log-partition-function, defined as $\Delta (\ln Z) \equiv \ln \hat{Z} - \ln Z = \ln \sum_i e^{- \beta \hat{E_i}} - \ln \sum_i e^{- \beta E_i}$. By assuming that the errors $e_i$ are smaller than the energies $E_i$, we can perform Taylor expansions on the log-sum-exp function $\ln \sum_i e^{- \beta \hat{E_i}} = \ln \sum_i e^{- \beta (E_i + e_i)}$ with respect to $e_i$. As shown in the Supplementary Note S3, after taking the expectation value with respect to the random variables $e_i$, the first-order term vanishes, and we have
\begin{align}
    \label{eqn:log_Z_second_order}
    \EX[\Delta (\ln Z)] \approx \frac{1}{2} \sigma^2 \beta^2 (1 - \alpha)
\end{align}
where $\alpha = \frac{\sum_i e^{-2 \beta E_i}}{(\sum_i e^{-\beta E_i})^2}$. Based on this result, according to Eq.~\eqref{eqn:thermodynamic}, the expected deviation of the configurational entropy due to the random energy deviations $e_i$, defined as $\EX[\Delta S] \equiv \EX[\hat{S} - S]$, is given by
\begin{align}
    \label{eqn: entropy}
    \EX[\Delta S] &= -\frac{1}{T} \frac{\partial}{\partial \beta} \EX[\Delta (\ln Z)] + k \EX[\Delta (\ln Z)] \nonumber \\
    &\approx -\frac{\sigma^2}{2kT^2} (1 - \alpha)
\end{align}
where we use the fact that $\alpha$, though containing $\beta$, does not depend on $\beta$ explicitly. Similarly, the expected deviation of the heat capacity is given by
\begin{align}
    \label{eqn: heat_capacity}
    \EX[\Delta C_v] &= \frac{1}{kT^2} \frac{\partial^2}{\partial \beta^2} \EX[\Delta (\ln Z)] \approx \frac{\sigma^2}{kT^2} (1 - \alpha)
\end{align}
The derivation details can be found in Supplementary Note S3. Therefore, both $\EX[\Delta S]$ and $\EX[\Delta C_v]$ are linearly dependent on the variance $\sigma^2$, but independent on the mean $\mu$, suggesting that the variance $\sigma^2$ of the energy deviations $e_i$ among the configurations must be minimized to accurately predict configurational entropy and heat capacity. 

To numerically verify this, we train our GNN model that is based on Transformer convolution layers on the pristine \AuCu~dataset, using the variance as the target loss function. The variance of the optimal model on the testing set is 15.82~meV\textsuperscript{2}/atom, but the MAE is 4.89~eV/atom; as a comparison, the variance of our previous model using MSE as the loss function is 13.2~meV\textsuperscript{2}/atom and the MAE loss is 2.76~meV/atom. In Fig.~\ref{fig:distortion_variance}(a), we show the comparison between the configurational energies predicted by GNN and DFT, suggesting a global shift of energies for all configurations. We then apply this model to MC simulations, and the predicted ordering-disorder phase transition temperature is 837~K, in good agreement with previous results using MSE as the loss function (870~K). Therefore, as long as the variance is minimized, the predicted configurational entropy and the heat capacity are reliable.

In conclusion, we demonstrate the capability of GNN with an attention mechanism to accurately predict the configurational disorder properties in compound materials, including configurational entropy and order-disorder phase transition temperature. Using the face-centered tetragonal \AuCu~dataset as an example, the predicted phase transition temperature from attention-based GNN models and MC simulations is close to that obtained in experiments. Even when random atomic displacements are introduced, reliable predictions are still achievable. Furthermore, we show that the variance of the configurational energy deviations between GNN and DFT controls the prediction accuracy of these disorder-related properties. These results provide new perspectives on the efficient and accurate evaluation of disordered properties in configurationally complex materials. This contributes significantly to future research focused on the phase stability of such materials and advances the exploration of high-entropy alloys and related material systems.

\section{Methods} \label{sec:methods}
\subsection{Density Functional Theory Calculations}
The first-principles calculations are performed based on DFT, as implemented in Vienna Ab initio Simulation Package \cite{Kresse93VASP1, Kresse96VASP2}. We use the projector augmented wave pseudopotentials \cite{Kresse99PAW1, Blochl94PAW2}, where $5d$ and $6s$ electrons are treated as valence electrons for Au and $3d$ and $4s$ electrons are treated as valence electrons for Cu. We used the GGA-PBE functional for all calculations \cite{Perdew96GGA} and 550~eV for the kinetic energy cutoff of the plane-wave basis sets. The $\bk$-point grid density is taken to be 0.03~$2 \pi$/\AA, and the energy convergence threshold is $10^{-7}$~eV.

After variable-cell relaxation, the ground state structure of stoichiometric AuCu is the face-centered-tetragonal structure. The relaxed lattice constants are $a = 2.86$~\AA~and $c = 3.55$~\AA. From the relaxed unit cell, we generate the $5 \times 5 \times 4$ supercell, such that the lattice constant along each direction is close to 15~\AA. Using the supercell, we generate the \AuCu~dataset covering the whole concentration range $0 < x < 1$.

\subsection{Constructing and Training GNN models}
In each layer of a GNN, the node features are updated by the features of neighboring nodes according to $\bx_i = \phi(\bx_i, \bx_{\mathcal{N}(i)})$, where $\bx_i$ is the feature of the $i$th node and $\mathcal{N}(i)$ is the adjacent nodes of $i$. This message aggregation process is repeated multiple times as the convolution layers are stacked, allowing GNNs to capture the long-range interactions in the crystal. Depending on the aggregate function $\phi$, various types of GNNs are proposed. In this work, we choose the attention-based Transformer network \cite{Shi21Transformer} to construct our GNN and generate the main results in the manuscript. 

For each layer, we use global mean pooling to extract the global graph feature from all nodes, as we choose the total energy per atom as the target quantity to predict. These global features are added together, forming shortcut connections that allow the gradient to flow more easily during training (benchmark results using only the global features from the last convolution layer can be found in the SM). Finally, the global features are passed to two fully-connected linear layers to predict the total energy per atom.

For training the GNN models, we shuffle the dataset and divide it into the training set, the validation set, and the testing set, allocating them in a 60:20:20 ratio. Unless otherwise stated, the target loss function is the MSE. The validation set is used to prevent overfitting on the training set. While training the GNN models, we keep track of the validation loss and use the model with the smallest validation loss as our final model.

To choose the optimal set of hyperparameters, we use the Bayesian optimization method as implemented in Optuna \cite{Optuna}, which calculates the expected improvement of the current set of hyperparameters based on results from previous trial runs using Tree-structured Parzen Estimator method \cite{Bergstra11TPE}. The convergence is achieved when no significantly different hyperparameter values are proposed for consecutive 10 trial runs. The hyperparameter set includes the number of convolution layers, the number of hidden channels, the number of attention heads (used only in GNN models with attention mechanisms), learning rate, weight decay, and batch size.

\subsection{Monte Carlo Simulations}
MC simulations with the Wang-Landau sampling method \cite{Wang01p2050} are then carried out to obtain the density of states $g(E)$, with the configurational energies evaluated from the trained GNN models. The flatness criterion is achieved when the minimum value of the histogram is no smaller than 80\% of the mean value, and the convergence is achieved when the modification factor satisfies $\ln{f} < 10^{-7}$. Since the configurational space of the supercell is complicated, at least around $10^8$ configurations are necessary for the MC simulation to converge, beyond the capabilities of first-principles methods. 

From the density of states, the configurational entropy and the heat capacity at an arbitrary temperature are calculated as 
\begin{align}
    \label{eqn:thermodynamic}
    C_v = \frac{1}{k_B T^2} (\langle E^2 \rangle - \langle E \rangle^2), \quad S = \frac{1}{T} (\langle E \rangle - F)
\end{align}
where the partition function is $Z = \int g(E) e^{-\beta E} dE$, the free energy is $F = - k_B T \ln{Z}$, the expected value for quantity $Q$ is $\langle Q \rangle = \frac{1}{Z} \int g(E) Q(E) e^{-\beta E} dE$, and $\beta = \frac{1}{k_B T}$.

\section*{Data Availability}
The dataset generated for this work are available on Github at \url{https://github.com/qmatyanlab/Configurational-Disorder.git}.

\section*{Code Availability}
The codes used in this work are available on Github at \url{https://github.com/qmatyanlab/Configurational-Disorder.git}.

\bibliography{bibliography}

\section*{Acknowledgements}
Z.F. thanks helpful advice from Weiyi Gong and Dr. Bojian Hou. Z.F. and Q.Y. acknowledge funding support from the U.S. Department of Energy, Office of Science, Basic Energy Sciences, under Award No. DE-SC0023664. This research used resources of the National Energy Research Scientific Computing Center (NERSC), a U.S. Department of Energy Office of Science User Facility located at Lawrence Berkeley National Laboratory, operated under Contract No. DE-AC02-05CH11231 using NERSC award BES-ERCAP0029544.

\section*{Author contributions statement}

Z.F. and Q.Y. conceived the experiments. Z.F. wrote the code for graph neural networks and Monte Carlo simulations and performed data analysis. Q.Y. supervised the study. The manuscript was written through contributions from all authors, and all authors have given approval to the final version of the manuscript.

\section*{Competing interests}
The authors declare no competing interests.

\section*{Additional information}
\subsection*{Supplementary Materials}
Supplementary Notes S1 to S3\\
Table S1\\
Figures S1 to S2\\

\end{document}